\documentclass[12pt]{iopart}

\expandafter\let\csname equation*\endcsname\relax
\expandafter\let\csname endequation*\endcsname\relax

\usepackage{graphicx}
\usepackage{caption}
\usepackage{xcolor}
\usepackage{subcaption}
\usepackage{tabularx}
\usepackage{hyperref}
\usepackage{amssymb}
\usepackage{physics}
\usepackage[utf8]{inputenc} 
\usepackage[nobreak]{cite}
\usepackage[T1]{fontenc}
\usepackage{siunitx}
\usepackage{upgreek}

\usepackage[noabbrev]{cleveref}
\usepackage{breakurl}

\crefname{equation}{}{}

\binoppenalty=\maxdimen
\relpenalty=\maxdimen
\hypersetup{
    colorlinks = true,
    citecolor={blue},
    linkcolor={red},
    allbordercolors = {white},
    urlcolor={blue}
}
 
\begin{document}

\title[All-fiber photon pair source for QKD in a network]{A flexible modular all-fiber based photon pair source for quantum key distribution in a network}

\author{Maximilian Tippmann, Erik Fitzke, Oleg Nikiforov, Philipp Kleinpa\ss, Till Dolejsky, Maximilian Mengler and Thomas Walther}

\address{Institute for Applied Physics, Technische Universität Darmstadt,\\Schlossgartenstra\ss e 7,  64289 Darmstadt, Germany}
\ead{thomas.walther@physik.tu-darmstadt.de}
\vspace{10pt}
\begin{indented}
\item[]\today
\end{indented}

\begin{abstract}
Entanglement-based QKD protocols require robust and stable photon pair sources in terms of high heralding efficiencies or photon pair generation rates even under harsh environmental conditions, e.g. when operated in the field. In this paper, we report on a flexible, tunable, alignment-free, all-fiber coupled photon source based on spontaneous parametric down-conversion in periodically poled crystals. It can be operated in continuous-wave and pulsed modes, respectively. Its rack-compatible and modular setup allows a straight forward plug-and-play integration of coding-modules e.g. interferometers to enable various QKD protocols such as phase or phase-time coding. We demonstrate operation as a type-II and a type-0 SPDC stage proving the high flexibility of our source. Furthermore, we demonstrate simultaneous operation of SHG and SPDC in a double-pass configuration within the same nonlinear crystal further simplifying the hardware requirements of our source. To evaluate the conversion efficiencies of our modules, we employ data post-processing to remove artefacts from detector afterpulsing and deadtimes of the detectors. We investigate the source performance for various repetition rates.
\end{abstract}

\vspace{2pc}

\maketitle

\section{Introduction}
Photon pair sources offer a large variety of applications such as quantum key distribution (QKD)~\cite{gisin2002, scarani2007, xu2020} and quantum computing \cite{kok2007}. Such sources can be used as heralded single photon sources \cite{montaut2017high, fasel2004, ngah2015} or to provide entangled pairs of photons \cite{stevenson2006, martin2010, lu2019, kwiat1995, steinlechner2012, anwar2021, cabrejo2022}. Each application comes with a unique set of source requirements in terms of photon production rates, noise suppression and filtering as well as photon separation and coupling for the experiments.

There are various methods to produce photon pairs. The most common is spontaneous parametric down-conversion (SPDC) typically in crystal waveguides employing periodic-poling for quasi-phase matching \cite{anwar2021}. However, for typical application scenarios in QKD setups the produced photon pairs are required to be coupled to an optical fiber to bridge larger distances. Thus, some photon pair sources use spontaneous four-wave-mixing in fibers to avoid coupling from a waveguide to a fiber reducing losses. These sources have the drawback to require cryogenic cooling of the setup to suppress Raman-noise photons which otherwise would degrade the signal-to-noise ratio of the source \cite{dyer2009, takesue2005}. 

With our source, we demonstrate that a good performance can be achieved with commercially available fiber-packaged crystal waveguides allowing a simple setup while accessing fibers to provide a stable source for entanglement-based QKD setups \cite{Four_party_system}. Many other sources are optimized towards a single purpose. For instance, these use asymmetric filtering in signal/idler paths, fixed pump powers, are not completely fiber-coupled or employ a seed laser at different wavelengths \cite{montaut2017high, ribordy2000, fiorentino2002, guo2017parametric}. However, our source can be used for various setups due to its plug-and-play design and high flexibility in terms of continuous wave (cw) to high repetition rate pulsed pumping for SPDC photon pairs. This allows an easy integration of coding modules such as interferometers for QKD protocols like phase-time-coding \cite{Four_party_system, marcikic2004} or adaption for phase-coding \cite{ribordy2000}. Producing photon pairs in the telecom C-band centered at 1550~nm, our source is ideally suited for use in real-world fiber QKD applications as this window offers lowest transmission losses in standard optical fibers. The source is completely fiber-coupled, hence dropping the need for any optical alignment.

\section{Source Setup}

\begin{figure*}[ht]
\includegraphics[width=0.95\textwidth]{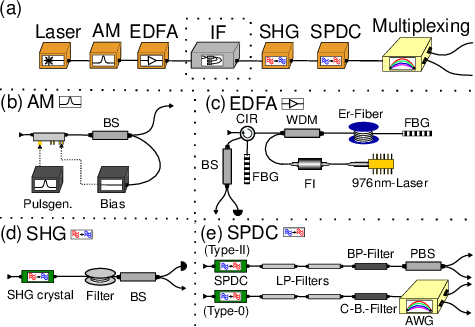}
\centering
\caption{Setup of the photon pair source. Figure (a) shows the complete source with its modules. The interferometer (IF) in the dashed box depicts the position where coding modules can be placed into the source setup. Figures (b)-(e) show the setup of each module in more detail. Note that there are two different setups for the SPDC-module, a type-II and a type-0 module. Abbreviations: AM - Amplitude Modulator; BS - Beamsplitter; FI - Faraday Isolator; FBG - Fiber Bragg Grating; Pulsgen. - Pulse Generator; Bias - Bias Controller; WDM - Wavelength Division Multiplexer; LP-Filter - Long Pass Filter; PBS - Polarization Beam Splitter; AWG - Arrayed Waveguide Grating; C-B.-Filter - C-Band Filter.}
\label{fig:setupsource}
\end{figure*}

The photon source features a modular design with six main components (cf. figure~\ref{fig:setupsource}(a)). The seed laser is a wavelength-stabilized DFB-laser (Wavelength References Clarity NLL-1550-HP) operating at 1550.51 nm on the edges of two adjacent 100~GHz ITU-DWDM grid channels. The laser is followed by an electro-optic modulator shaping pulses of flexible duration and repetition rate from the cw radiation of the seed laser. Then, the ensuing laser pulses are amplified employing an in-house made erbium-doped fiber amplifier (EDFA). This is followed by type-0 second harmonic generation (SHG) in a periodically poled LiNbO$_3$ crystal. In the final step, we use type-II or type-0 spontaneous parametric down-conversion (SPDC) in a periodically poled LiNbO$_3$ crystal waveguide  as well. Throughout the setup, several beamsplitters/tap couplers are placed used either for diagnostic purposes or as feedback for power stabilization. Depending on the intended use of the photon pairs we introduce an imbalanced interferometer (IF) between the EDFA and the SHG step which acts as a coding module for QKD applications e.g. employing a phase-time protocol~\cite{Four_party_system}. Notably, by starting with 1550~nm radiation,  undergoing SHG followed by SPDC we obtain several advantages: First, commercially available components for telecom wavelengths can be used, hence dropping the need of expensive TiSa lasers at 775~nm. Second, by placing the interferometer in front of the SHG it operates at 1550~nm. Thus, it can be manufactured by the same method and with the same components as receiver interferometers for a phase-time coding setup like \cite{Four_party_system}, drastically simplifying the precise fabrication. This proves especially useful in these kinds of setups as all interferometers from source and receivers have to be matched in their arm lengths with a precision of a few hundred micrometers \cite{Four_party_system}. 

The electro-optic modulator shapes pulses followed by a beamsplitter where a small portion (10~\%) of the light is fed to the bias controller. This device delivers a DC offset to the electro-optic modulator to either lock the electro-optic modulator to minimum or maximum transmission. A pulse generator (either HP8131A or HP8133A) generates the electrical pulses to switch the electro-optical modulator to produce optical pulses. If cw operation is required, the pulse generator can be switched off and the bias controller is set to maximum transmission locking. Hence, no physical change of the setup is required to switch from pulsed to cw operation.
\begin{figure}[ht]
\includegraphics[width=0.8\columnwidth]{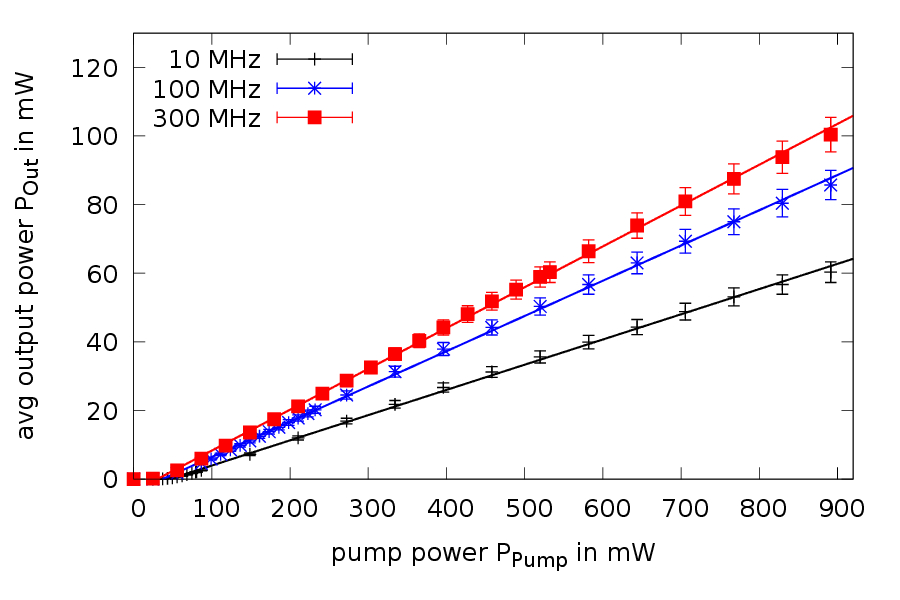}
\centering
\caption{Averaged EDFA output power for pulsed operation at different pump power and pulse repetition rates. The pulse shapes are nearly rectangular with estimated FWHM pulse widths of 433~ps for a repetition rate of 10~MHz, 396~ps for 100~MHz and 501~ps for 300~MHz, respectively.}
\label{fig:edfa}
\end{figure}

The in-house built double-pass EDFA incorporates fiber Bragg gratings serving as pump and amplified spontaneous emission (ASE) filters achieving a spectrally pure seed output with residual light suppressed by at least 80~dB. The pump laser can deliver a power of up to 900~mW at 976~nm. This leads to a large variety of accessible output powers for various repetition rates and pulse lengths. Thus, average output powers of up to 100~mW at 300~MHz pulse repetition rate with pulse durations of 501~ps have been observed (cf. figure~\ref{fig:edfa}).

\begin{figure}[ht]
\includegraphics[width=0.8\columnwidth]{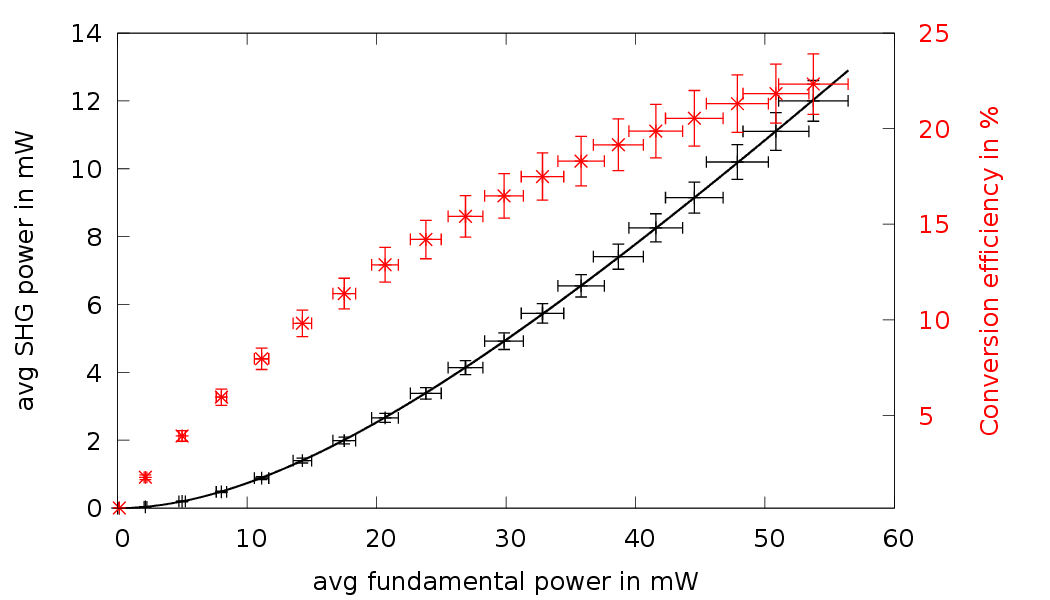}
\centering
\caption{Averaged output powers of the SHG crystal when pumped with fundamental pulses of 501~ps FWHM at 300~MHz repetition rate. The SHG-power is measured at the 99~\% output of the tap coupler (cf. figure~\ref{fig:setupsource}(d)).}
\label{fig:shg}
\end{figure}

For the SHG stage, the setup uses 5 meters of tightly coiled PM780 fiber following the SHG crystal to filter out residual fundamental light at 1550~nm. When tested with 2~meters of fiber, again, very high suppression of at least 79~dB is reached without causing significant insertion loss ($<1~$dB) for the SHG light at 775~nm. The SHG crystal is a 34~mm long PPLN Type-0 crystal waveguide from NTT Electronics. The crystal features conversion efficiencies of up to 22~\% at only 54~mW averaged fundamental power (equivalent to approximately 360~mW pulse peak power) without using a resonator setup (cf. figure~\ref{fig:shg}). The pulse duty cycle is approximately 15~\% which allows SHG pulse peak powers of around 100~mW, when increasing the pump even further. Of course, for photon pair generation aiming for QKD only fundamental powers of less than 10 mW are necessary. 

The curve of the generated SHG power can be explained considering a simple pump depletion model following \cite{Kiefer_2019} and extending it by a constant to factor in the coupling loss at the crystal's facets,
\begin{equation}
    P_{\textrm{SHG}} = c^2P_{\textrm{Fundamental}}\cdot \tanh^2\left(\sqrt{\eta_{\textrm{BK}}\cdot c \cdot P_{\textrm{Fundamental}}}\right)
\end{equation}
with the coupling ratio $c$, the Boyd-Kleinman conversion efficiency $\eta_{\textrm{BK}}$, and the pulse peak power $P_{\textrm{Fundamental}}$ calculated from the average fundamental power $\bar P$, the pulse repetition rate $r$ and pulse length $\tau$ according to:
\begin{equation}
    P_{\textrm{Fundamental}} = \frac{\bar P}{r \cdot \tau} \textrm{ .}
\end{equation}
For simplicity of the fit model the coupling ratio is set to be equal for both facets.  As demonstrated, our setup enables a wide range of available SHG powers to ensure a large accessible region of different mean photon pair numbers per pulse $\mu$ at the SPDC process.

Each SPDC module consists of a PPLN-crystal for photon pair generation followed by two long pass filters to remove the remaining 775~nm light. The subsequent 7.1~nm band pass (full-width half-maximum) or C-band filter (cf. figure~\ref{fig:setupsource}(e)) makes sure the photon pairs are within the operation ranges of the photon separation filters to avoid undesired crosstalk between the channels. To separate the photons, we employed a polarization beam splitter (PBS) for the type-II process or a standard telecom C-band arrayed-waveguide-grating (AWG) with 100~GHz channel separation for the type-0 process, respectively. A comparison of both crystals or modules can be found in table \ref{tab_SPDC}. The crystals are usually operated at $43.56~^\circ$C in case of the type-0 and $41.64~^\circ$C for the type-II crystal, respectively, via package integrated thermo-electric coolers. The temperature is stabilized within a 10 mK range of these temperatures.

The complete source is made of fiber-coupled components with the crystals being fiber coupled and packaged as well. The modules are connected via standard optical fibers. These connections allow flexible source configuration e.g. bypassing of the amplitude modulator or integrating interferometers but introduce additional losses because of the fiber connectors. However, this does not degrade the performance of the source as all losses before the SPDC process can be simply mitigated by increasing the pump power of the EDFA. The setup is polarization maintaining until photon separation to ensure a stable pump power at the SPDC crystals in the relevant polarization axis, thus allowing stable photon pair generation rates. Further stabilization of pump power can be added by adapting the driver current of the EDFA via a software-based PID loop using the power measured at the beamsplitter after SHG as feedback. If additional pump power is needed, the plug-and-play design allows an easy integration of further EDFAs.

\begin{table*}
	\centering
\caption{\label{tab_SPDC}Comparison of type-0 and type-II SPDC crystals (above horizontal line) and modules (below horizontal line).}
\begin{indented}
\item[]\begin{tabular}{@{}lll}
 \br
		 Property & Type-0 & Type-II \\
		 \mr
		 manufacturer & NTT Electronics & AdVR\\
		 material & ZnO:PPLN & MgO:PPLN \\
		 crystal length & 34~mm & 24~mm  \\
		 fiber coupled & yes & yes \\
		 est. conversion efficiency & \((4.8\pm 0.2)\times10^{-7}\)& \((7.6\pm 0.4)\times10^{-10}\)\\
		 & (within 7.1~nm band)& \\
		 & 5.1~THz usable width with & \\
		 &C-band filter&\\
		 \mr
		 Filters & longpass & longpass   \\
		 & + C-Band filter &+ 7.1~nm bandpass\\
		 Photon pair separation & wavelength  & polarization \\
		 & (symmetric to center WL)& (slow/fast) \\
 \br
	\end{tabular}
\end{indented}
\end{table*}

\subsection{Double-pass type-0 configuration}

\begin{figure*}[tb]
\includegraphics[width=0.95\textwidth]{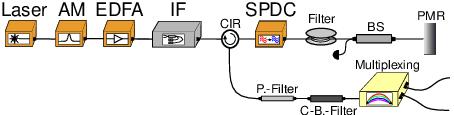}
\centering
\caption{Setup for photon pair production in double-pass configuration. The type-0 SPDC module is used to produce SHG light in the forward direction and photon pairs in the reverse direction. Abbreviations: PMR - Polarization Maintaining Retroreflector; P.Filter - Pump light Filter; C-B.-Filter - C-Band Filter.}
\label{fig:doublepass}
\end{figure*}

Both, the SHG crystal and the type-0 SPDC crystal are 34-mm long PPLN-Crystals. If the same crystal is used for both SHG and SPDC, the complexity and cost of the photon pair source is considerably reduced. Cascaded SHG and SPDC within a single crystal have been demonstrated~\cite{arahira2011generation} utilizing a single-pass configuration. However, this can make filtering of noise photons tedious. While there are setups utilizing a double-pass configuration with a single crystal, these are usually designed such that polarization entangled photons can be produced~\cite{anwar2021, kim2019pulsed, steinlechner2013phase}. As our source does not produce polarization entanglement, we have investigated an operation mode of the source where the type-0 SPDC crystal was used to produce SHG light in forward direction and photon pairs in backward direction. The setup is shown in figure~\ref{fig:doublepass}.

To operate the photon pair source with a single crystal a polarization maintaining circulator is placed in front of the type-0 module (without the C-band filter), sending 1550-nm pump light through the long pass filters into the crystal. 
After passing a wound fiber the produced 775-nm light is retro-reflected back into the crystal. 
As in the SHG-module a 90:10 beam splitter is introduced to monitor the back-propagating SHG power that now pumps the SPDC process.  
Finally, the generated photon pairs are guided through a pump light filter followed by the C-band filter before they are distributed via wavelength-division demultiplexing. 
The pump light filter is used to remove any remaining laser light and consists of two DWDM filters and two fiber Bragg gratings.
This filter as well as all other added optical components are connected through standard mating-sleeves, thus rebuilding the source from single-pass to double-pass configuration only requires reconnecting optical fibers.

\section{Experimental Details and Results}

\subsection{Heralding Efficiency}

An important feature of photon sources is the heralding efficiency, i.e. the probability that an idler photon is extracted when the signal has been detected~\cite{pittman2005heralding}.
We estimated the heralding efficiency of our source for both modules and for different channel combinations using the setup depicted in figure \ref{fig:setupsource}(a). Both outputs (for the type-0 module two frequency conjugate outputs of the AWG) are connected to a detection unit each. The photons are detected with two single-photon avalanche detectors (ID Quantique ID220) with set detection efficiencies of 10~\% and dead times of 5~$\mu$s. The detection events are time-tagged by an ID Quantique ID900 time controller. A pulse repetition rate of 33~MHz was chosen.
The heralding efficiency for channel 1 and 2 is given by \cite{meyer2017limits} $\eta_{1/2} = C_{12}/(\eta_{\textrm{Det}} R_{2/1} )$ with the count rates in each channel $R_i$, the coincidence rate between both channels $C_{12}$ and the detector efficiency $\eta_{\textrm{Det}}$ to account for the losses due to imperfect photon detection. The actual efficiency for our detectors is 10.6~\% and 9.3~\%~\cite{Fitzke_2022}. The type-II module allows heralding efficiencies of up to 36~\% at 1.8~pJ per pump pulse. The heralding efficiencies of the type-0 module are lower, reaching up to 11.3~\% at best for the channel at 191.7~THz while ranging between 8.7~\% and 12.3~\% for the other tested DWDM channels at approximately 0.05~pJ per pump pulse. The pump powers were chosen to achieve similar count rates at the detectors for the type-II and type-0 process. The results for all channels are displayed in figure \ref{fig:herald}. The large difference between the type-II and type-0 modules can be attributed to the different multiplexing technique accompanied by a higher insertion loss of the AWG. Thus, the heralding efficiency can be significantly improved by using low-loss filters and gratings for photon separation. For the type-II module, we chose a symmetric filter configuration in contrast to \cite{montaut2017high} allowing both fast and slow axis to achieve similar heralding efficiencies e.g. enabling both axes for use as signal/herald arm. This is important as the source is featured as a photon pair source and not solely a heralded single photon source. We also note that the heralding efficiencies for both modules can be improved by directly splicing the SPDC-crystals' fibers to the subsequent filters instead of using standard fiber connectors. The same holds for the C-band filter in the type-0 module. 
The count rates yielded from the heralding efficiency measurements can be utilized to estimate the source brightness following the method from \cite{montaut2017high} with 
\begin{equation}\label{eq:brightness}
    B = \frac{R_1 R_2}{C_{12}P_{\textrm{Pump}} t \Delta \lambda}   \textrm{ .}
\end{equation}
This yields brightnesses of $2.7 \times 10^6$ pairs/(s$\cdot$mW$\cdot$nm) for the type-II module (at 1.8~pJ per pulse) and $3.9 \times 10^8$ to $4.4 \times 10^8$ pairs/(s$\cdot$mW$\cdot$nm) (at $0.048-0.056$~pJ per pulse) for the type-0 module, respectively. The values obtained via~\eref{eq:brightness} do not take into account count rate / coincidental rate distortions induced by the detector dead times. This might be negligible for detectors with short dead times compared to the count rates per second such as superconducting-nanowire detectors \cite{idq}. For detectors where this is not the case, one detector being in dead time will cause an underestimation of the coincidental rates leading to an overestimation of the brightness.

\subsection{Conversion Efficiencies}

\begin{figure*}[tb]
\includegraphics[width=0.95\textwidth]{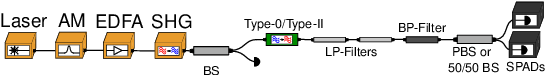}
\centering
\caption{Setup to measure the photon pair production rates and estimate the respective conversion efficiencies of both (type-0/type-II) waveguides. Only one SPDC waveguide is inserted into the measurement setup at a time. Following the filters, the setup is connected to two single-photon avalanche detectors (ID Quantique ID220). Note that for the type-0 waveguide a 7.1~nm-band pass filter was used instead of the C-band filter to allow a better comparison to the type-II process in terms of conversion efficiency. The detection events are time-tagged by an ID Quantique ID900 (not depicted). For the type-0 process a 50/50 beamsplitter is used instead of the polarization beam splitter.}
\label{fig:setupconveff}
\end{figure*}

To overcome this issue of imperfect detectors the brightness can be determined by evaluating the crystals' conversion efficiencies with a slightly more sophisticated model. To estimate the conversion efficiencies or the photon pair production rates of our SPDC crystals we employed a setup shown in figure~\ref{fig:setupconveff}. The SHG module was followed by a 90:10 beam splitter to monitor the SHG power impinging on the SPDC crystals. Either one of the type-0 or type-II crystals had been connected to the 90~\% output of the beam splitter and was followed by the long pass filters to remove the SHG light. Contrary to the setup of the SPDC modules in figure~\ref{fig:setupsource}(e) a 7.1~nm band pass filter has been placed into the setup for both types of crystals to allow for a better comparison of the respective conversion efficiencies. 

For both crystals several measurements at various input (SHG) powers have been conducted. From the timestamps of both detectors a photon pair production rate can be deduced applying equations for the coincidence and count rate statistics following~\cite{hayat1999theory} to
\begin{equation}\label{eq:coneff}
    \mu_\textrm{gen} = \gamma\frac{\eta_\textrm{pair}}{\eta_s\eta_i}\frac{\left( \frac{N_1}{T} - r_1 \right)\left( \frac{N_2}{T} -r_2 \right) T}{N_\mu}\textrm{ .}
\end{equation}
$N_i$ is the number of photon counts detected at a single detector $i$ within measurement time $T$. The quantity $r_i$ is the measured dark count rate of a detector. $N_\mu$ is the number of coincidence counts without accidental counts. 
The factor $\gamma$ in~\eref{eq:coneff} is $\gamma = 1$ for deterministic photon pair splitting, e.g. using a PBS for the type-II process, or $\gamma = 1/2$ for probabilistic splitting, e.g. using a 50/50 beam splitter in case of the type-0 process. The factor 1/2 comes from the 50/50 beam splitter guiding both photons into the same channel with 50\% probability, so that the measured coincidence rate is halved. 
The factors $\eta$ arise from the fact that the losses are not frequency-independent for photon pair production i.e. $\eta_\textrm{pair} \neq \eta_i \eta_s$.
Due to the spectral correlation of the signal and idler photons, the detection probabilities for signal and idler photons are not independent. 
Hence, these factors are defined via the given filters' transmission functions $p(f)$ as
\begin{equation}
    \eta_S= \frac{1}{\Delta F} \int_F p_S(\nu_0+f)df \textrm{ and } \eta_I= \frac{1}{\Delta F} \int_F p_I(\nu_0-f)df
    \label{eq:si}
\end{equation}
\begin{equation}
    \eta_\textrm{pair}= \frac{1}{\Delta F} \int_F p_S(\nu_0+f)p_I(\nu_0+f)df
    \label{eq:pair}
\end{equation}
where $F$ is a frequency interval with width $\Delta F$.

\begin{figure}[bt]
\includegraphics[width=0.97\columnwidth]{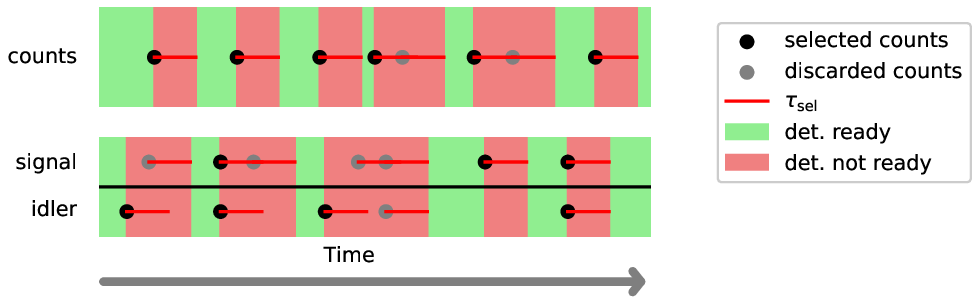}
\centering
\caption{Post-selection of the crystal efficiency measurement data to remove afterpulsing and dead time effects. Upper diagram: Correction for individual count rates. Lower diagram: Correction for coincidence rates. If an event has been registered the detector is considered to be not ready for time $\tau_\textrm{sel}$ thus ignoring subsequent events within this timespan. For the coincidence correction a subsequent event is kept if it occurs within the coincidental time window.}
\label{fig:postselection}
\end{figure}

However, as shown in~\cite{Fitzke_2022} the detectors are prone to afterpulsing and dead time effects. One detector being in dead time while the other is not, will lead to an underestimation of the ratio between coincidences to single counts. Afterpulsing manifests itself as a detection event causing additional clicks after the dead time ended although no photon was incident on the detector. For the type-II measurement the detectors showed a maximum afterpulsing probability of 5.11~\% and 3.43~\% respectively. In addition, this considerably distorts the coincidental to single count rate ratio, hence corrupting the estimate of the conversion efficiencies. In order to avoid this detection drawback, post-processing the recorded timestamps to remove such effects from the data has been introduced~\cite{rapp2019dead, coates1968correction}. As found in~\cite{Fitzke_2022} the afterpulsing is limited to several tens of $\mu s$ after the dead time ended. By choosing an extended dead time of $\tau_\textrm{sel} = 40~\mu s$ we post-select the data as depicted in figure~\ref{fig:postselection}. 

\begin{figure}[tb]
\centering
\begin{subfigure}{0.49\columnwidth}
\includegraphics[width=\textwidth]{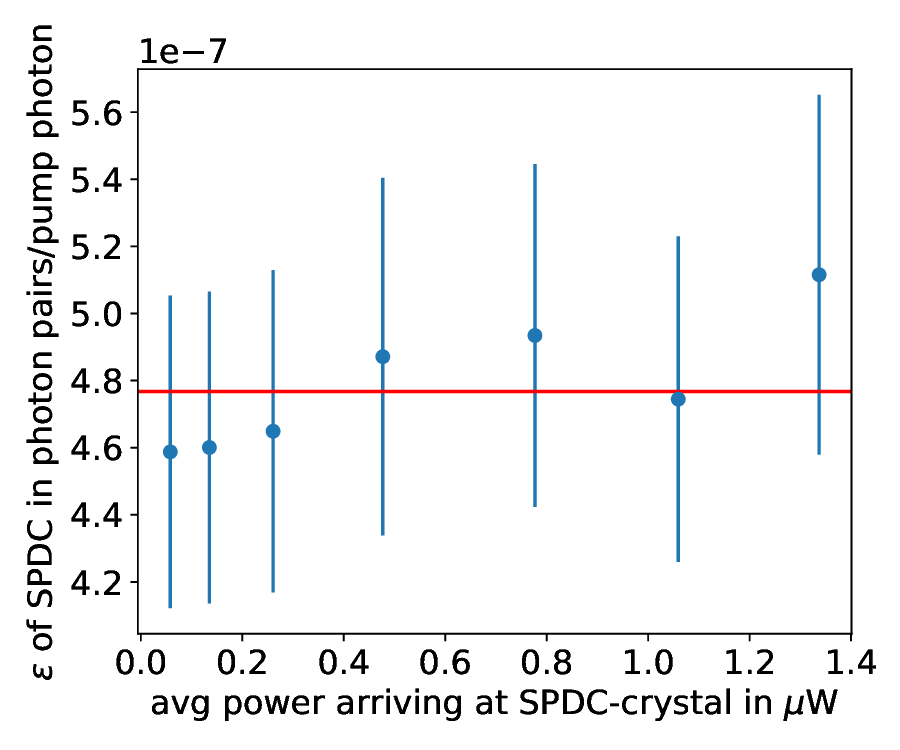}
\caption{}
\end{subfigure}
\hfill
\begin{subfigure}{0.49\columnwidth}
\includegraphics[width=\textwidth]{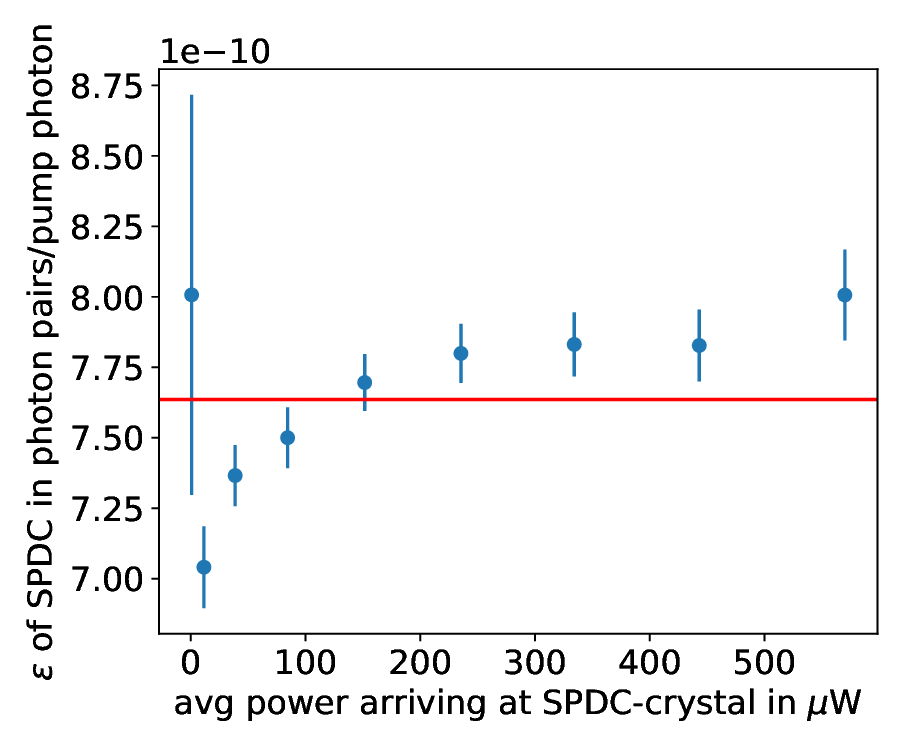}
\caption{}
\end{subfigure}
\caption{SPDC conversion efficiency for various SHG powers impinging on the type-0 waveguide (a) and type-II waveguide (b). The horizontal lines indicate the weighted average conversion efficiencies of the crystals. The large error bar for the first data point in (b) is due to low count rates at the respective pump power.}
\label{fig:conveff}
\end{figure}

In general, a detection event is only considered when outside the detectors deadtime, i.e. when the event was detected after a sufficiently long period of time after the previous event. To evaluate the single count rate of a detector all timestamps within $\tau_\textrm{sel}$ after a previous detection are removed. For coincidences both detectors have to be considered: Here, $\tau_\textrm{sel}$ is applied to both detectors, although an event only has been detected at one detector. However, if the second event is within the coincidence time of the previous event, it is still kept as a valid detection. To yield count rates, the number of counts and coincidences of the post-selected data are divided by the total time a detector has been ready.  
Using our previous results for the photon pair production rates and the post-selected data we have calculated conversion efficiencies for every pump power. The results are depicted in figure~\ref{fig:conveff}. An average weighted with the uncertainty of the data points is estimated to state the respective conversion efficiencies. The weighted average is indicated by a horizontal line in figure~\ref{fig:conveff}. This corresponds to conversion efficiencies of $(4.8\pm 0.2)\times 10^{-7}$ for the type-0 and $(7.6\pm0.4 )\times 10^{-10}$ for the type-II crystal in terms of photon pairs produced per impinging photon at 775~nm. For the type-0 crystal the conversion efficiency is estimated using the 7.1-nm band pass filter only cutting off large parts of the spectrum, as can be seen from the spectrum in the following section. However, this method is convenient as it allows a better comparison to the type-II process and avoids distortion of the measurement introduced by components (e.g. 50/50 coupler) for wavelengths far from the center wavelength. Nevertheless, the conversion efficiency of the type-0 crystal is significantly higher than the estimated value when considering a larger wavelength range.

The brightness of our source can be directly calculated from the conversion efficiencies when considering the pump photon flux and the respective spectral widths. For the type-0 spectrum a brightness of $B= 3.32 \times 10^{8} \textrm{~photon pairs}/(\textrm{s nm mW})$ is achieved. The type-II module shows a brightness of $B= 2.48 \times 10^6 \textrm{~photon pairs}/(\textrm{s nm mW})$ when taking into account that the full-width-half-maximum of the spectra is estimated to 1.2~nm (cf. figure~\ref{fig:type2spectra}), i.e. much smaller than the 7.1~nm transmission window of the installed filter. 
Both values are in close vicinity of the previously estimated values from~\eref{eq:brightness}, confirming the applicability of our model. 

For QKD applications the photon pairs will be separated and sent via different channels to several parties, hence a good estimate of the mean photon number produced per pump pulse $\mu$ per channel is required to control parameters like time-base QBER e.g. in phase-time protocols avoiding multiple photon pairs per detection cycle~\cite{Four_party_system}. For the type-II process, we can separate the photons via polarization into two channels, so the mean photon number per pump pulse per channel can be calculated by 
\begin{equation}
    \mu = \eta_{\textrm{conv}} \cdot N_{\textrm{Pump}}
\end{equation} 
with the respective process' conversion efficiency $\eta_{\textrm{conv}}$ and the number of pump photons per pulse $N_{\textrm{Pump}}$. For the type-0 process, the formula has to be adapted, as the photon separation occurs via wavelength multiplexing instead of polarization. Hence, the mean photon number per wavelength channel has to be considered which strongly depends on transmission characteristics of the installed photon separation device e.g. an AWG. Thus, it is more convenient to state a spectral conversion efficiency which is $\eta_{\nu}=6.81\times 10^{-7} /\textrm{THz}$ and calculating the mean photon number from it employing the respective channel width. For further evaluation of the photon transmissions it is important to take into account the spectral dependence of the respective wavelength-multiplexer employing~\eref{eq:si} and~\eref{eq:pair}.

\subsection{Photon Pair spectra}

\begin{figure*}[tb]
\includegraphics[width=0.95\textwidth]{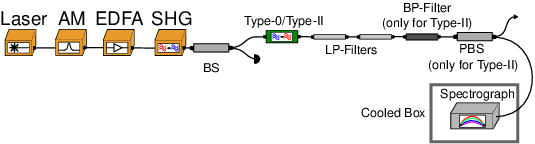}
\centering
\caption{Setup for measurement of the photon pair spectra. Only one SPDC waveguide is inserted into the measurement setup at a time. For the type-0 waveguide, no band pass filter was used allowing measurement of the full spectral width of the generated photon pairs. Furthermore, a polarization beam splitter is not required for the type-0 spectrum measurements. The setup uses an Andor Shamrock 500i Czerny-Turner spectrograph with a 600 lines/mm grating, 1900~nm blaze and an Andor iDus InGaAs $1.7~\mu$m CCD detector. The spectrograph setup is placed inside a box cooled down to close to $0~^\circ$C in order to further lower the temperature of the camera chip and allow for single-photon detection.}
\label{fig:setupspectra}
\end{figure*}

\begin{figure}[t]
\centering\includegraphics[width=0.95\columnwidth]{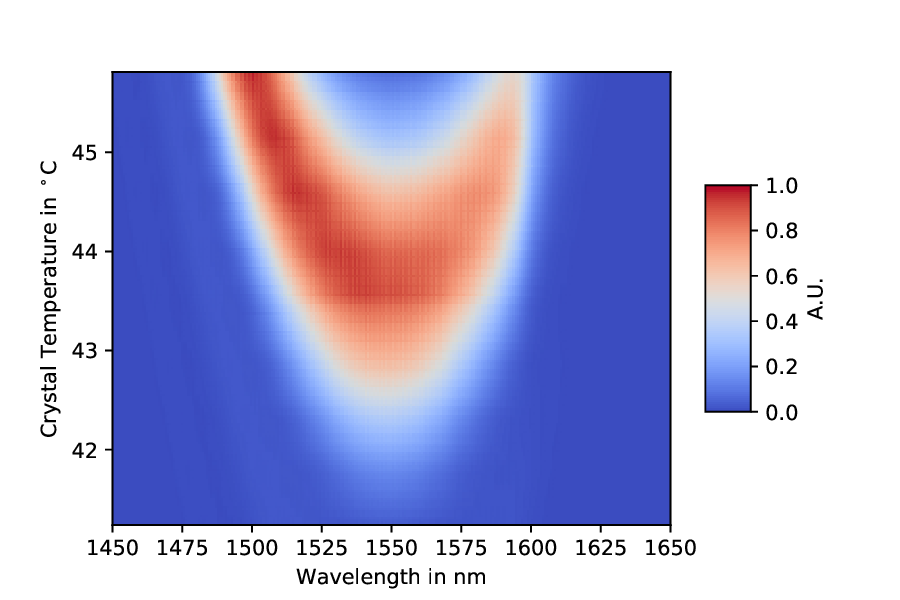}
\caption{Photon pair spectra of the type-0 waveguide for various waveguide temperatures. The spectra were linearly interpolated and displayed with Gouraud shading for better visibility. Towards higher temperatures one can see a broadening of the spectrum, while for lower temperatures the spectrum narrows and the photon pair production rates are reduced. The tendency to higher intensity towards the left half of the figure is attributed to a slightly reduced detection efficiency of our spectrograph for higher wavelengths. At temperatures of around 44~$^\circ$C the spectrum is basically flat and has a width of around 100~nm. Employing an arrayed waveguide grating, this wide spectrum enables simultaneous QKD with dozens of receivers.}
\label{fig:type0spectra}
\end{figure}

\begin{figure}[ht]
\centering\includegraphics[width=0.95\columnwidth]{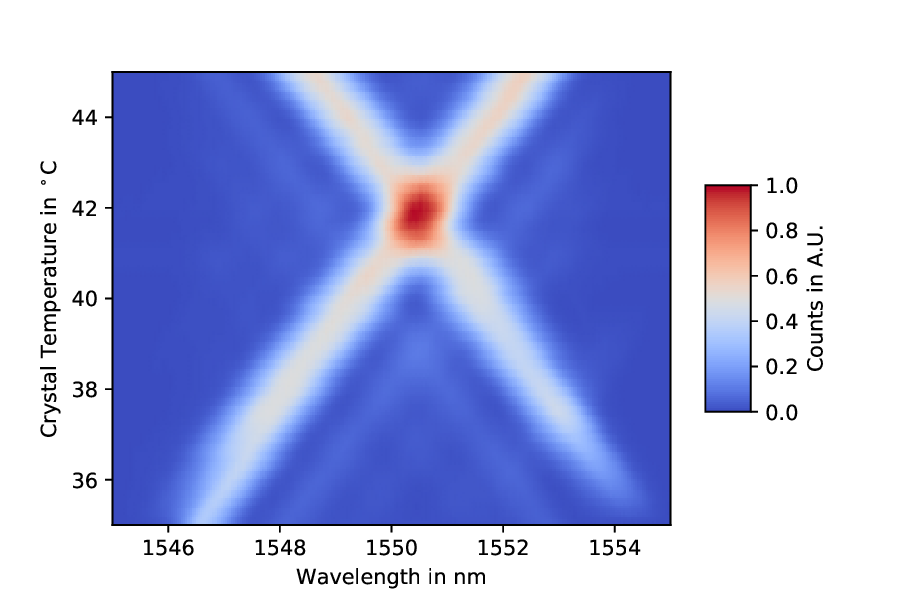}
\caption{Photon pair spectra of the type-II waveguide for various waveguide temperatures. Depicted is the sum of the fast and slow axis outputs. The spectra were linearly interpolated and displayed with Gouraud shading for better visibility. When tuning the temperature of the waveguide, one can clearly observe the expected X-shaped structure, given by a wavelength degenerate state at approximately $42~^\circ$C combined with the crossover for both polarization states at this point. The doubling of the intensity at the intersection compared to the exterior regions arises from adding the rates of the two polarization states generated in the type-II process. The relatively low intensities on the red side of the spectrum at $35~^\circ$C are due to the transmission edge of the 7.1~nm band pass filter.}
\label{fig:type2spectra}
\end{figure}

As a next step, we investigated the spectra for both SPDC crystals for various waveguide temperatures. Using the setup depicted in figure~\ref{fig:setupspectra} each temperature setting has been measured separately. The results have been merged into figures~\ref{fig:type0spectra} and \ref{fig:type2spectra} for better comparison of the tested temperatures. The  operating temperatures were chosen in such a way that the type-II process generates degenerate photons while the type-0 process offers a wide, almost flat spectrum in a region of high conversion efficiencies, allowing efficient, controllable QKD operation as well as many accessible ITU-DWDM channels. The latter case is especially important if the source shall be used in a setting with several simultaneous pairwise key distribution as in~\cite{Four_party_system}.

\begin{figure}[tb]
\centering
\begin{subfigure}{0.49\textwidth}
    \includegraphics[width=\textwidth]{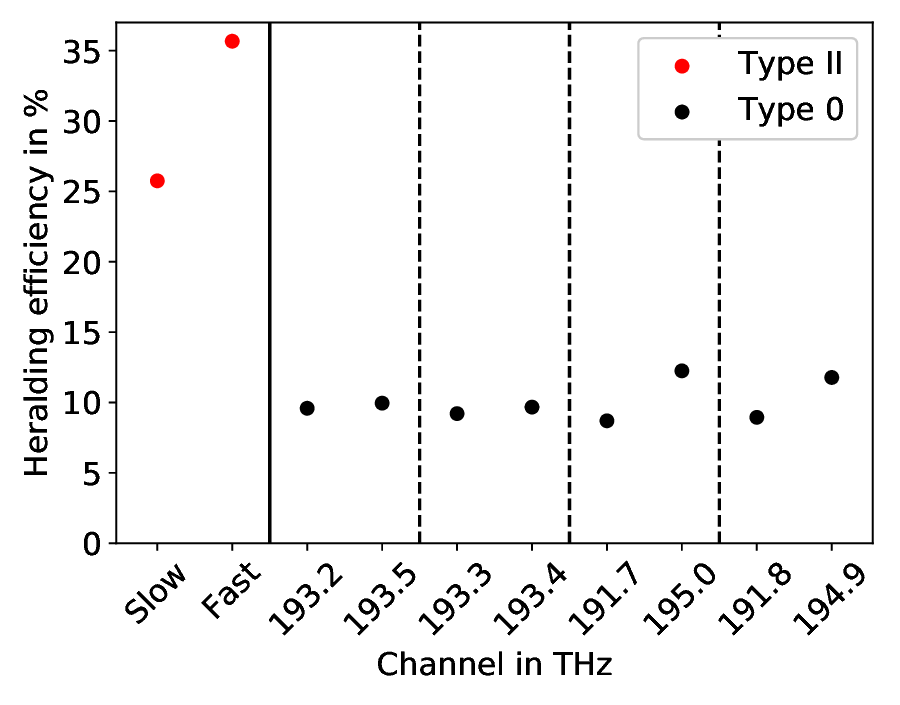}
    \caption{\label{fig:herald}}
\end{subfigure}
\hfill
\begin{subfigure}{0.49\textwidth}
    \includegraphics[width=\textwidth]{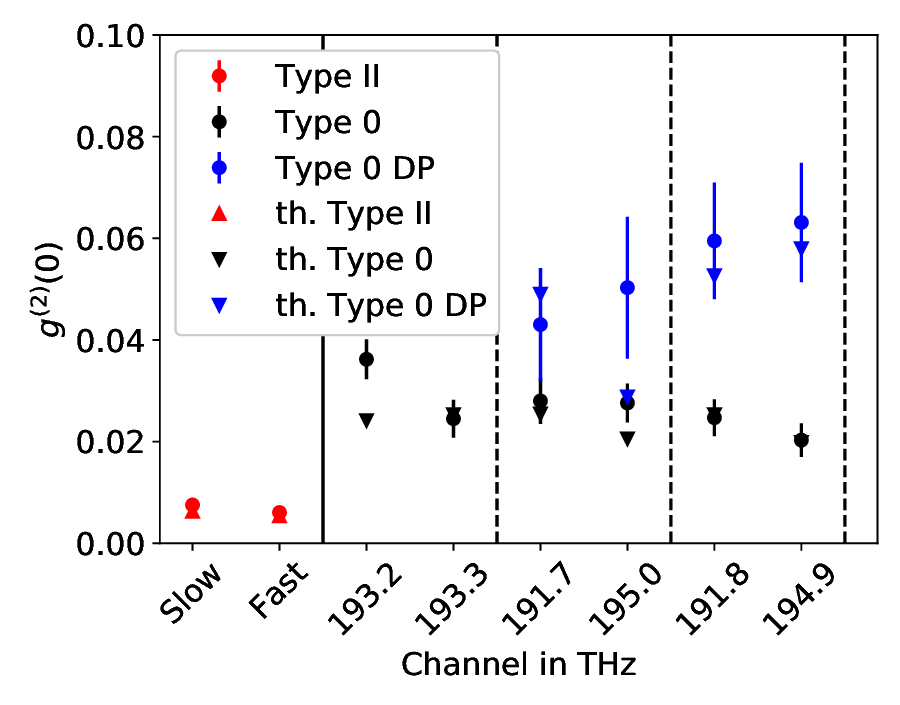}
    \caption{\label{fig:g2}}
\end{subfigure}
\caption{(\subref{fig:herald}) displays the heralding efficiencies for various tested channels for both kinds of SPDC processes. (\subref{fig:g2}) shows the obtained values for the second-order Glauber-correlation. Error-bars for the type-II process are not displayed for visibility, as they are much smaller than the type-0 errors. Triangles depict the theoretical $g^2(0)$ values based on multi-pair events.}
\label{fig:photonstats}
\end{figure}

To characterize the quantum character of our source, we have estimated the heralded second-order Glauber correlation function $g^{(2)}_h(0)$ for both SPDC processes. This value is a measure for the bunching character of our source, which is $g^{(2)}_h(0) = 0 $ for an ideal single photon source/heralded photon pair source. Values larger than zero indicate photon pair production with multi-photon events or noise e.g. from non-correlated background photons.
SPDC is a statistical process sometimes producing multiple photon pairs per pulse.
The larger the pump power of the SPDC process the higher the share of pulses containing multiple photon pairs after the SPDC. As a consequence the choice of the pump power directly affects the minimal $g^{(2)}(0)$ value a source can achieve. 
To setup the measurement of $g^{(2)}_h(0)$ one output of the photon pair source is directly connected to a detector. The other output is connected to a 50/50 beam splitter with both of these outputs being routed to a detector each. In case of the type-II process the outputs are the orthogonal polarization axes, while for the type-0 process pairs of channels with entangled wavelengths have been connected. Measuring the dark count corrected count and coincidental rates, $R_i$ and $C_{ijk}$, $g^{(2)}_h(0)$ can be obtained by
\begin{equation}
    g^{(2)}_h(0)= \frac{C_{123}R_3}{C_{13}C_{23}}
\end{equation}
where $i=3$ indicates the heralding arm (detector being directly connected to output)~\cite{beck2007comparing}. This gives $g^{(2)}_h(0) \approx (6\pm0.8)\times 10^{-3}$ for the fast axis of the type-II process and $(7.5\pm0.9)\times 10^{-3}$ for the slow axis (pulse energy 0.91~pJ and 1.06~pJ respectively, equivalent to $\mu = 2.7 \times 10^{-3}$ to $3.1\times 10^{-3}$). Both values are close to $g^{(2)}_h(0)=0$ displaying the good single photon character of our source.
For the type-0 process several channel combinations with pulse energies ranging from $4.3 \times 10^{-2}$~pJ to $5.5\times 10^{-2}$~pJ ($\mu = 1.5\times 10^{-2}$ to $1.9\times 10^{-2}$) have been measured (cf. figure~\ref{fig:photonstats}\subref{fig:g2}). The observed $g^{(2)}_h(0)$ ranges from $2.0 \times 10^{-2}$ to $3.6\times 10^{-2}$. We expect a large proportion of this value to be given by multi-photon emissions which are dependent on the pulse energy. As an estimate we calculate the $g^{(2)}_h(0)$ for a Poissonian distribution of multi-pair events. Considering one-photon-pair-events as the main contribution for dual coincidences and two-photon-pair-events for triple coincidences while neglecting higher order terms,
\begin{equation}
    g^{(2)}(0)= \frac{P(I_1, I_2, S)P(S)}{P(I_1,S)P(I_2,S)}
    = \mu\left( 2\frac{\eta_S \eta_I}{\eta_\textrm{pair}} - \eta_S T_S\right) \textrm{ .}
\end{equation}
While $T_S$ is the transmission probability of a signal photon, the factors $\eta$ are defined by transmission functions of the AWG $p(f)$ following~\eref{eq:si} and~\eref{eq:pair}. Consequently, $\eta_S \eta_I/\eta_\textrm{pair}$ gives a factor of approximately 0.67 for a frequency interval of 130~GHz. Taking into account the various losses on the path for a signal photon (crystal decoupling, filter transmissions, detector efficiency, it follows that $2\eta_S \eta_I/\eta_\textrm{pair} \gg \eta_S T_S$, hence
\begin{equation}
   g^{(2)}_h(0)\approx 2\frac{\eta_S \eta_I}{\eta_\textrm{pair}}\mu \textrm{ .}
\end{equation}
Employing the spectral $\mu$ density, found from the estimated conversion efficiency, for 130~GHz one finds the estimates for $g^{(2)}_h(0)$ at our pump energies. The values are depicted in figure~\ref{fig:photonstats}\subref{fig:g2}. We observe a good agreement of our measured $g^{(2)}_h(0)$ values with the value predicted by the number of multi-pair events. Hence, the emission characteristics of our source can be mostly explained by considering multi-pair emissions.

The measurement was repeated with the photon source in double-pass configuration (cf. figure~\ref{fig:doublepass}). Although the $g^{(2)}_h(0)$ value is independent of losses, the additional losses introduced by the circulator and the pump-filter required a higher pump power in order to obtain suitable count statistics. This resulted in a higher $\mu$-value of around $4\times 10^{-2}$ (with the exception for the 195.0~THz measurement at $\mu\approx 2.1\times 10^{-2}$) and therefore yielded higher $g^{(2)}_h(0)$ values between $4.3\times 10^{-2}$ and $6.3\times 10^{-2}$.
For both configurations, all values are close to zero and way below the standard threshold for considering a source as a heralded single photon source of $g^{(2)}_h(0)=0.5$. Again, there is a good agreement with the theoretical predictions for multi-pair emissions, when taking into account the measurement uncertainties. 
 
\begin{figure*}[ht]
\centering\includegraphics[width=0.95\textwidth]{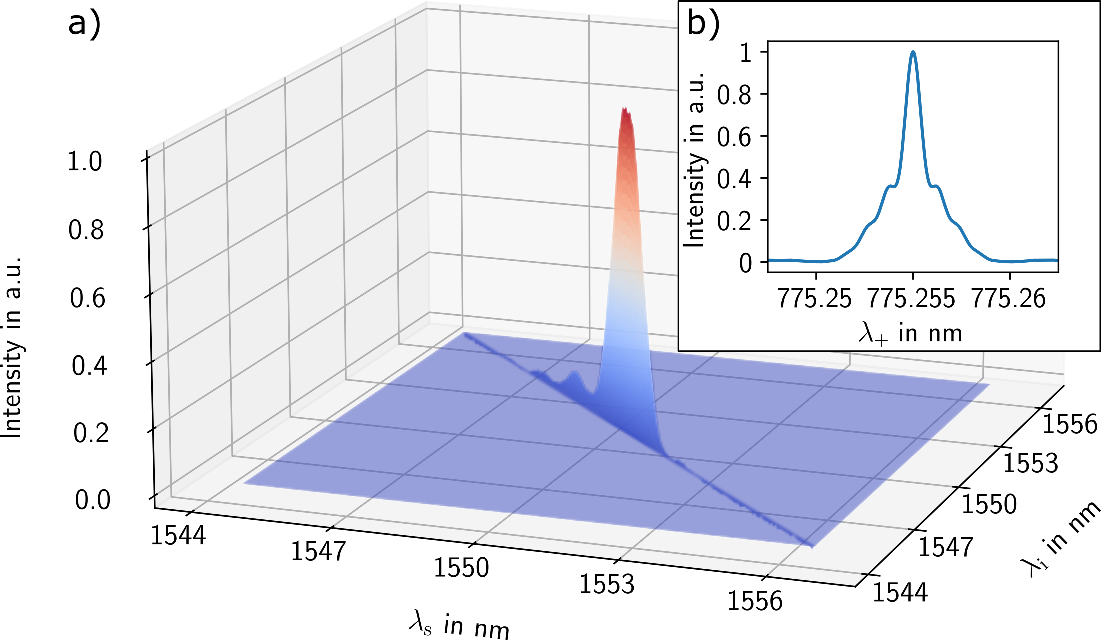}
\caption{(a) Joint spectral intensity for type-II SPDC at the degeneracy temperature, pumped by a transform-limited pulse of approximately $\SI{400}{\pico\second}$ duration. The pump spectrum was obtained by fast Fourier transform of a temporal pulse shape measurement and is presented in (b), where $\lambda_+ = 2 \pi c/\omega_+$. The corresponding Schmidt number amounts to $K \approx 120$, classifying the generated state as highly entangled with a large number of independent squeezers.
}
\label{fig:type2jsi}
\end{figure*}

To approximate the photon emission statistics of a photon pair source, we consider the spectra of the photon pairs generated by the SPDC process. These are determined by the joint spectral amplitude (JSA) 
\begin{equation}
\psi(\omega_{\mathrm{s}}, \omega_{\mathrm{i}}) = \alpha(\omega_{\mathrm{s}} + \omega_{\mathrm{i}}) \Phi(\omega_{\mathrm{s}}, \omega_{\mathrm{i}}),
\end{equation}
where $\alpha$ describes the pump pulse distribution and $\Phi$ is the phase-matching function of the nonlinear crystal and $\omega_{\mathrm{s}}, \omega_{\mathrm{i}}$ are the angular frequencies of the signal and idler photons with wavelengths $\lambda_{\mathrm{s}}, \lambda_{\mathrm{i}}$. Even for pump pulse durations as short as $\SI{400}{\pico\second}$, the bandwidth of the photon pair spectra is approximately two orders of magnitude wider than the pump pulse spectrum. Thus, we may assume that the phase-matching function is constant over the narrow range of $\omega_+ = \omega_{\mathrm{s}} + \omega_{\mathrm{i}}$ governed by the pump pulse and solely depends on the difference frequency $\omega_- = \omega_{\mathrm{s}} - \omega_{\mathrm{i}}$. This allows us to reconstruct the JSA from measurements of the pump pulse shape and the SPDC spectra acquired by cw-pumping. The resulting joint spectral density $|\psi(\omega_{\mathrm{s}}, \omega_{\mathrm{i}})|^2$ is presented in figure \ref{fig:type2jsi} a) for the type-II crystal at the degeneracy temperature and a pump duration of approximately $\SI{400}{\pico\second}$, for which the corresponding spectrum is given in figure \ref{fig:type2jsi} b).
The shape of the JSA in anti-diagonal $\lambda_- = 2\pi c/\omega_-$ direction is determined by the phase matching function. For an ideal crystal with length~$L$ and phase mismatch $\Delta k = [n(\omega_+)\omega_+ - n(\omega_s)\omega_s-n(\omega_i)\omega_i]/c$, the phase matching function is given by $\Phi(\omega_{\mathrm{s}}, \omega_{\mathrm{i}}) = \mathrm{sinc}^2[\Delta k L /2]$~\cite{Ou_2007}. Figure~\ref{fig:type2jsi}~b) shows different sizes of the side lobes at both sides of the central wavelength. Although side lobes are expected from the sinc\textsuperscript{2}-function, its symmetry is not met, most likely due to imperfections of the crystal. 

An important measure to quantify the entanglement between the signal and idler photons is given by the Schmidt number $K = 1 / \sum_j \lambda_j^2$, where the Schmidt coefficients $\lambda$ are given by the expansion coefficients of the Schmidt decomposition \begin{equation}
\psi(\omega_{\mathrm{s}}, \omega_{\mathrm{i}}) = \sum_j \sqrt{\lambda_j} u_j(\omega_{\mathrm{s}}) v_j(\omega_{\mathrm{i}})
\end{equation}
with sets of orthogonal function~$\{u(\omega_s)\}$ and $\{v(\omega_i)\}$~\cite{Schmidt_1907, Law_2000, Lamata_2005, Mauerer_2009}. 

Calculating the Schmidt-decomposition of the discretized joint spectral amplitude of the type-II process presented in figure \ref{fig:type2jsi} via singular value decomposition~\cite{Lamata_2005,Mauerer_2009_PhDThesis,Christ_2011,Christ_2013}
yields a Schmidt number of $K \approx 120$. For the type-0 process even larger Schmidt numbers $K \gg 100$ are obtained, classifying the two-photon states produced by our photon source as highly entangled states compared to the regimes considered for example in \cite{Mauerer_2009, Christ_2011}.
The unheralded second-order Glauber correlation function can be approximated from the number of Schmidt modes via $g_u^{(2)}(0)=1+ \frac{1}{K}$~\cite{Christ_2011, Eckstein_2011}. A thermal photon-number distribution for a single two-mode squeezer, resembled by $K=1$, gives $g_u^{(2)}(0)=2$, while for an infinite number of two mode squeezers it converges to a Poissonian photon number distribution. Hence, small Schmidt numbers indicate a thermal photon distribution, while large Schmidt numbers display a Poissonian photon emission~\cite{Christ_2011, Mauerer_2009}. Following $g_u^{(2)}(0)=1+ \frac{1}{K}$, we obtain $g_u^{(2)}(0) = 1 + \frac{1}{120}\approx 1.008 $ which underlines the Poissonian nature ($g_u^{(2)}(0)=1)$ of our SPDC source.

\noindent 

\section{Discussion and Outlook}
We demonstrated a versatile alignment-free modular photon-pair source being able to host plug-and-play type-0 and type-II SPDC modules.
Intrinsically, the photon separation of the type-II process via polarization allows to connect two receivers for the photons. If one aims for larger QKD networks to demonstrate simultaneous pairwise key exchange with multiple pairs of receivers  \cite{Wengerowsky2018, Four_party_system}, this can be achieved by adding wavelength-division multiplexing filters at each polarization channel. Another variant would be to employ beamsplitters (e.g. 1x2 50:50 splitters) and implement time-division multiplexing~\cite{frohlich2013quantum}. Of course, the latter comes with a trade-off in terms of key rates when considering pairwise links as the routing of the photons is not deterministic anymore.  

The type-0 process with its wide spectrum is ideally suited for such types of QKD networks, allowing many pairs of receivers being connected to a single photon pair source. The AWG used in our setup in combination with the C-band filter allows the use of 17 ITU-channel pairs (34 channels of 100~GHz). As the C-band filter offers a usable range of $\pm2.55$~THz around the center frequency of our photon pair spectrum one could drastically increase the number up to 102 users by employing a 50~GHz channel spacing AWG while accessing the complete usable C-band filter range. The type-0 spectra is approximately 9.3~THz wide. Hence, replacing the C-band filter can increase the number of WDM-connected users even further.

The source has been tested for various operational modes from cw to pulsed operation with several hundred of MHz repetition rates and pulses as short as 400~ps. In principle, even higher repetition rates up to a few GHz and pulse durations down to 150~ps should be applicable as the amplitude modulator is specified for frequencies up to 10~GHz. Hence, largely variable photon pair production rates can be achieved. This is an asset, as there are many types of single photon detectors with different detection efficiencies and dead times. If the photon pair production rate does not match the detector properties and experimental details, such as transmission losses e.g. in fiber QKD experiments, the detector can be driven into saturation, thus being in dead time for a large proportion of the measurement time. Considering entanglement-based experiments this will cause a drop of the coincidence rates compared to the count rates of the detectors, considerably reducing the signal-to-noise ratio. This effect can be mitigated when adapting the photon pair production rate with respect to the capabilities of the detectors.

Furthermore, the large tunability of our source is especially useful for QKD protocols with interferometers such as phase-time coding. Here, the repetition rate and pulse duration must be chosen with respect to the delay introduced by the interferometers' arm length differences to avoid pulse overlapping. Consequently, our source can be employed in settings with various interferometers without the need to change any hardware.

The EDFA and SHG modules generate pump pulses without background from amplified spontaneous emission at a large range of powers enabling a wide range of accessible mean photon numbers per pulse $\mu$ in the SPDC module. Tuning $\mu$ can be important in various ways. First, it can affect the detector saturation as described for the photon pair production rate above. Second, in QKD protocols like phase-time coding it does yield a minimum bound for the quantum bit error rate (QBER) of the setup. For our source, as the Schmidt numbers prove, the SPDC is a Poissonian process. Hence, the number of events where not only a single photon pair but multiple photon pairs are emitted from a single pump pulse increases with higher $\mu$. These extra photon pairs will cause additional clicks in the detectors not being correlated to the clicks caused by the original photon pair, thus increase the QBER. Here, the Poissonian nature of our source is advantageous, as the probability for multi-pair events at low $\mu$ is lower than for a thermal distribution, as can be shown by evaluating the multi-pair emission probabilities for both types of photon number distributions. Consequently, by tuning $\mu$ our source does allow to choose between higher raw key rates vs lower QBER, depending on the applicable scenario in such QKD systems. 

When considering the performance of our source it is important to stress that the crystals are of-the-shelf components with no particular coupling optimization. In terms of performance (e.g. heralding efficiencies, in-source losses of photon pairs) all losses occurring after photon pair production are particularly costly. Thus, an optimized coupling of the SPDC crystals could yield a better source performance. However, this would require to step back from standard products. As an easier point of improvement, we can take into account any fiber-to-fiber connectors. These typically account for 0.3~dB of loss each. We employed such connectors to attach the SPDC crystal's output to the subsequent filters. In case of the type-0 module the C-band filter introduces another fiber connector. This accounts to a total of $1.2~\textrm{dB} = 2 \times 0.6~\textrm{dB}$ of avoidable loss for the type-0 module when considering that both photons of a pair experience these losses. To lift the attenuation one could replace the named connection by fiber splices which hardly cause losses for telecom wavelengths standard fibers.

\section{Conclusion}
We report on two flexible single photon-pair sources. The system consists of spectrally pure generation of 775~nm light, allowing subsequent SPDC modules to generate photon pairs around 1550~nm. The first is based on a type-II SPDC process and the second on a type-0 SPDC process. Their robust design makes them ideal for many applications either as a heralded single-photon source or as a photon pair source as basis e.g. for entanglement based QKD protocols. Both can be operated either in cw or pulsed operation. 
Furthermore, we have demonstrated that the source can produce photon pairs of high quality with a single crystal operated in a double-pass configuration for both SHG and SPDC. In case of cw operation, the source is ideally suited for phase-coding QKD applications \cite{ribordy2000}. In pulsed operation, we have demonstrated a small QKD network based on time-bin entanglement~\cite{Four_party_system}. We report on a brightness of $2.48\times 10^6~\textrm{photon pairs}/(\textrm{s nm mW})$ for the type-II process and $3.32\times 10^8~\textrm{photon pairs}/\textrm{s nm mW}$ for the type-0 process, with heralding efficiencies up to 36~\% and from 9~\% to 12~\%, respectively.
The heralded Glauber correlation function $g^{(2)}(0)$ takes values of $<7.5\times 10^{-3}$ and $<3.6\times 10^{-2}$ for $\mu$ of $\approx 3.1\times 10^{-3}$ and  $\approx 1.9\times 10^{-2}$, respectively, showcasing a good photon pair emission behaviour.

\section*{Acknowledgements}
This research has been funded by the Deutsche Forschungsgemeinschaft (DFG, German Research Foundation), under Grant No. SFB 1119--236615297. We thank Paul Wagner from Deutsche Telekom Technik GmbH for lending us the AWG.

\section*{References}
\providecommand{\noopsort}[1]{}\providecommand{\singleletter}[1]{#1}%
\providecommand{\newblock}{}

\clearpage


\begin{thebibliography}{10}
\expandafter\ifx\csname url\endcsname\relax
  \def\url#1{{\tt #1}}\fi
\expandafter\ifx\csname urlprefix\endcsname\relax\def\urlprefix{URL }\fi
\providecommand{\eprint}[2][]{\url{#2}}

\bibitem{gisin2002}
Gisin N, Ribordy G, Tittel W and Zbinden H 2002 {\em Rev. Mod. Phys.\/} {\bf
  74}(1) 145--195

\bibitem{scarani2007}
Scarani V, Bechmann-Pasquinucci H, Cerf N~J, Du\ifmmode~\check{s}\else
  \v{s}\fi{}ek M, L\"utkenhaus N and Peev M 2009 {\em Rev. Mod. Phys.\/} {\bf
  81}(3) 1301--1350

\bibitem{xu2020}
Xu F, Ma X, Zhang Q, Lo H~K and Pan J~W 2020 {\em Rev. Mod. Phys.\/} {\bf
  92}(2) 025002

\bibitem{kok2007}
Kok P, Munro W~J, Nemoto K, Ralph T~C, Dowling J~P and Milburn G~J 2007 {\em
  Rev. Mod. Phys.\/} {\bf 79}(1) 135--174
  \urlprefix\url{https://link.aps.org/doi/10.1103/RevModPhys.79.135}

\bibitem{montaut2017high}
Montaut N, Sansoni L, Meyer-Scott E, Ricken R, Quiring V, Herrmann H and
  Silberhorn C 2017 {\em Physical Review Applied\/} {\bf 8} 024021

\bibitem{fasel2004}
Fasel S, Alibart O, Tanzilli S, Baldi P, Beveratos A, Gisin N and Zbinden H
  2004 {\em New Journal of Physics\/} {\bf 6} 163

\bibitem{ngah2015}
Ngah L~A, Alibart O, Labont{\'e} L, d'Auria V and Tanzilli S 2015 {\em Laser \&
  Photonics Reviews\/} {\bf 9} L1--L5

\bibitem{stevenson2006}
Stevenson R~M, Young R~J, Atkinson P, Cooper K, Ritchie D~A and Shields A~J
  2006 {\em Nature\/} {\bf 439} 179--182

\bibitem{martin2010}
Martin A, Issautier A, Herrmann H, Sohler W, Ostrowsky D~B, Alibart O and
  Tanzilli S 2010 {\em New Journal of Physics\/} {\bf 12} 103005

\bibitem{lu2019}
Lu X, Li Q, Westly D~A, Moille G, Singh A, Anant V and Srinivasan K 2019 {\em
  Nature physics\/} {\bf 15} 373--381

\bibitem{kwiat1995}
Kwiat P~G, Mattle K, Weinfurter H, Zeilinger A, Sergienko A~V and Shih Y 1995
  {\em Phys. Rev. Lett.\/} {\bf 75}(24) 4337--4341
  \urlprefix\url{https://link.aps.org/doi/10.1103/PhysRevLett.75.4337}

\bibitem{steinlechner2012}
Steinlechner F, Trojek P, Jofre M, Weier H, Perez D, Jennewein T, Ursin R,
  Rarity J, Mitchell M~W, Torres J~P {\em et~al.\/} 2012 {\em Optics express\/}
  {\bf 20} 9640--9649

\bibitem{anwar2021}
Anwar A, Perumangatt C, Steinlechner F, Jennewein T and Ling A 2021 {\em Review
  of Scientific Instruments\/} {\bf 92} 041101

\bibitem{cabrejo2022}
Cabrejo-Ponce M, Spiess C, Muniz A~L~M, Ancsin P and Steinlechner F 2022 {\em
  Quantum Science and Technology\/} {\bf 7} 045022

\bibitem{dyer2009}
Dyer S~D, Baek B and Nam S~W 2009 {\em Optics express\/} {\bf 17} 10290--10297

\bibitem{takesue2005}
Takesue H and Inoue K 2005 {\em Optics express\/} {\bf 13} 7832--7839

\bibitem{Four_party_system}
Fitzke E, Bialowons L, Dolejsky T, Tippmann M, Nikiforov O, Walther T, Wissel F
  and Gunkel M 2022 {\em PRX Quantum\/} {\bf 3}(2) 020341
  \urlprefix\url{https://link.aps.org/doi/10.1103/PRXQuantum.3.020341}

\bibitem{ribordy2000}
Ribordy G, Brendel J, Gautier J~D, Gisin N and Zbinden H 2000 {\em Physical
  Review A\/} {\bf 63} 012309

\bibitem{fiorentino2002}
Fiorentino M, Voss P~L, Sharping J~E and Kumar P 2002 {\em IEEE Photonics
  Technology Letters\/} {\bf 14} 983--985

\bibitem{guo2017parametric}
Guo X, Zou C~l, Schuck C, Jung H, Cheng R and Tang H~X 2017 {\em Light: Science
  \& Applications\/} {\bf 6} e16249--e16249

\bibitem{marcikic2004}
Marcikic I, De~Riedmatten H, Tittel W, Zbinden H, Legr{\'e} M and Gisin N 2004
  {\em Physical Review Letters\/} {\bf 93} 180502

\bibitem{Kiefer_2019}
Kiefer D~C 2020 {\em Ultraviolette Laser zur K{\"u}hlung relativistischer
  Ionenstrahlen\/} Ph.D. thesis Technische Universit{\"a}t Darmstadt
  \urlprefix\url{http://tuprints.ulb.tu-darmstadt.de/11312/}

\bibitem{arahira2011generation}
Arahira S, Namekata N, Kishimoto T, Yaegashi H and Inoue S 2011 {\em Optics
  express\/} {\bf 19} 16032--16043

\bibitem{kim2019pulsed}
Kim H, Kwon O and Moon H~S 2019 {\em Scientific reports\/} {\bf 9} 1--7

\bibitem{steinlechner2013phase}
Steinlechner F, Ramelow S, Jofre M, Gilaberte M, Jennewein T, Torres J~P,
  Mitchell M~W and Pruneri V 2013 {\em Optics express\/} {\bf 21} 11943--11951

\bibitem{pittman2005heralding}
Pittman T, Jacobs B and Franson J 2005 {\em Optics communications\/} {\bf 246}
  545--550

\bibitem{meyer2017limits}
Meyer-Scott E, Montaut N, Tiedau J, Sansoni L, Herrmann H, Bartley T~J and
  Silberhorn C 2017 {\em Physical Review A\/} {\bf 95} 061803

\bibitem{Fitzke_2022}
Fitzke E, Krebs R, Haase T, Mengler M, Alber G and Walther T 2022 {\em New
  Journal of Physics\/}
  \urlprefix\url{https://doi.org/10.1088/1367-2630/ac5004}

\bibitem{idq}
{ID Quantique SA} 2021 "{I}{D} 281 {S}uperconducting nanowire system -
  {P}roduct brochure"
  \urlprefix\url{https://marketing.idquantique.com/acton/attachment/11868/f-023b/1/-/-/-/-/ID281_Brochure.pdf}

\bibitem{hayat1999theory}
Hayat M~M, Torres S~N and Pedrotti L~M 1999 {\em Optics communications\/} {\bf
  169} 275--287

\bibitem{rapp2019dead}
Rapp J, Ma Y, Dawson R~M and Goyal V~K 2019 Dead time compensation for
  high-flux depth imaging {\em ICASSP 2019-2019 IEEE International Conference
  on Acoustics, Speech and Signal Processing (ICASSP)\/} (IEEE) pp 7805--7809

\bibitem{coates1968correction}
Coates P 1968 {\em Journal of Physics E: Scientific Instruments\/} {\bf 1} 878

\bibitem{beck2007comparing}
Beck M 2007 {\em JOSA B\/} {\bf 24} 2972--2978

\bibitem{Ou_2007}
Ou Z~Y~J 2007 {\em Multi-Photon Quantum Interference -\/} (Berlin Heidelberg:
  Springer Science \& Business Media) ISBN 978-0-387-25554-5

\bibitem{Schmidt_1907}
Schmidt E 1907 {\em Mathematische Annalen\/} {\bf 63} 433--476

\bibitem{Law_2000}
Law C~K, Walmsley I~A and Eberly J~H 2000 {\em Phys. Rev. Lett.\/} {\bf 84}(23)
  5304--5307

\bibitem{Lamata_2005}
Lamata L and León J 2005 {\em Journal of Optics B: Quantum and Semiclassical
  Optics\/} {\bf 7} 224–229 ISSN 1741-3575

\bibitem{Mauerer_2009}
Mauerer W, Avenhaus M, Helwig W and Silberhorn C 2009 {\em Phys. Rev. A\/} {\bf
  80}(5) 053815

\bibitem{Mauerer_2009_PhDThesis}
Mauerer W 2009 {\em On Colours, Keys, and Correlations: Multimode Parametric
  Downconversion in the Photon Number Basis\/} doctoralthesis
  Friedrich-Alexander-Universit{\"a}t Erlangen-N{\"u}rnberg (FAU)

\bibitem{Christ_2011}
Christ A, Laiho K, Eckstein A, Cassemiro K~N and Silberhorn C 2011 {\em New
  Journal of Physics\/} {\bf 13} 033027
  \urlprefix\url{https://doi.org/10.1088/1367-2630/13/3/033027}

\bibitem{Christ_2013}
Christ A, Brecht B, Mauerer W and Silberhorn C 2013 {\em New Journal of
  Physics\/} {\bf 15} 053038
  \urlprefix\url{https://dx.doi.org/10.1088/1367-2630/15/5/053038}

\bibitem{Eckstein_2011}
Eckstein A, Christ A, Mosley P~J and Silberhorn C 2011 {\em Phys. Rev. Lett.\/}
  {\bf 106}(1) 013603
  \urlprefix\url{https://link.aps.org/doi/10.1103/PhysRevLett.106.013603}

\bibitem{Wengerowsky2018}
Wengerowsky S, Joshi S~K, Steinlechner F, H{\"{u}}bel H and Ursin R 2018 {\em
  Nature\/} {\bf 564} 225--228 ISSN 14764687 (\textit{Preprint}
  \eprint{1801.06194})
  \urlprefix\url{http://dx.doi.org/10.1038/s41586-018-0766-y}

\bibitem{frohlich2013quantum}
Fr{\"o}hlich B, Dynes J~F, Lucamarini M, Sharpe A~W, Yuan Z and Shields A~J
  2013 {\em Nature\/} {\bf 501} 69--72

\end{thebibliography}
\end{document}